\documentclass{iau_FM}
\usepackage{graphicx}
\usepackage{amsmath}
\usepackage{amssymb}
\usepackage{float}

\newcommand{\enzo}{\it{\small ENZO}}

\title[Spectral analysis of simulated galaxy clusters] 
{Spectral analysis of magnetic fields in simulated galaxy clusters}

\author[P. Dom{\'i}nguez-Fern{\'a}ndez, F. Vazza \&M. Br{\"u}ggen]   
{Paola Dom\'inguez-Fern\'andez $^1$, Franco Vazza$^{3,1,2}$
 \and Marcus Br{\"u}ggen$^1$}

\affiliation{$^1$Hamburger Sternwarte, Gojenbergsweg 112, 21029 Hamburg, Germany \\ email: {\tt pdominguez@hs.uni-hamburg.de} \\email: {\tt mbrueggen@hs.uni-hamburg.de}\\[\affilskip]
$^2$  Istituto di Radio Astronomia, INAF, Via Gobetti 101, 40121 Bologna, Italy\\[\affilskip]
$^3$  Dipartimento di Fisica e Astronomia, Universit\'{a} di Bologna, Via Gobetti 92/3, 40121, Bologna, Italy \\email: {\tt franco.vazza2@unibo.it}
}

\pubyear{2018}
\jname{Astronomy in Focus} 
\editors{Teresa Lago, ed.}
\begin{document}

\maketitle

\begin{abstract}

We introduce a new sample of galaxy clusters obtained from a cosmological simulation covering
an unprecedented dynamical range. All the clusters in our sample show a clear signature of small-scale dynamo amplification.
We show that it is possible to use dynamo theory 
for studying the magnetic spectrum in the intracluster medium.
We study if the intrinsic variations on the spectra depend on the dynamical history of each cluster
or on some host cluster properties.

\end{abstract}

\section{Introduction}

Galaxy clusters  evolve 
mainly via two mechanisms: accretion of gas and galaxies, and via mergers 
occurring approximately every few Gyr. These mechanisms directly 
affect the diffuse, hot, weakly magnetised gas observed in clusters referred as 
the Intracluster Medium (ICM) (e.g. \cite{Kravtsov}, \cite{Brunetti}) by
driving shocks through it and therefore, every merger event can be a source of turbulence. 
In particular, these events effectively affect the structure of the
magnetic fields that permeate galaxy clusters.

We know from radio observations that the strength of magnetic fields in the ICM
is of the order of  $\mu$G and the coherence scales are of the order of 10-50 kpc 
(e.g. \cite{Feretti,Govoni}). These 
coherent large-scale fields indicate that the mechanism which leads to this final 
structure cannot be solely due to gas compression. In fact, a small-scale dynamo
would be a more suitable mechanism to generate observed,
large-scale magnetic fields. While the origin of magnetic fields in the Universe remains
 an open question, a small-scale dynamo seems to play a key role in amplifying 
magnetic fields in galaxy clusters (e.g. \cite{Beres}), independent of where the seed fields come from.

\section{Methods and Results}

We studied the formation of seven massive galaxy clusters in a cosmological MHD simulation with the {\enzo} grid code (\cite{enzo})
using the Dedner cleaning method, and  
eight adaptive mesh refinement (AMR) levels to increase the dynamical resolution. 
Each cluster was selected in a  comoving volume of (260 Mpc)$^3$ and further refined in most of the volume in which the clusters form 
with a maximum spatial resolution of $\Delta x_{\rm max}=3.95 ~\rm kpc$ (comoving).  We assumed a simple uniform 
seed field of cosmological origin of 0.1 nG (comoving) at $z=30$ (for a detailed description see \cite{Vazza18}). In this work, we will 
only discuss the results from {\it non-radiative} cosmological simulations to focus on the growth 
of magnetic fields by the turbulence induced by structure formation processes.
 
\begin{figure}[H]
\centering
	\includegraphics[width=7.5cm]{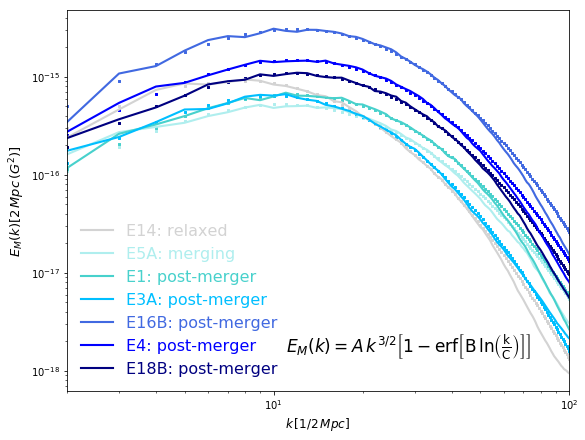}
	\includegraphics[width=6.8cm]{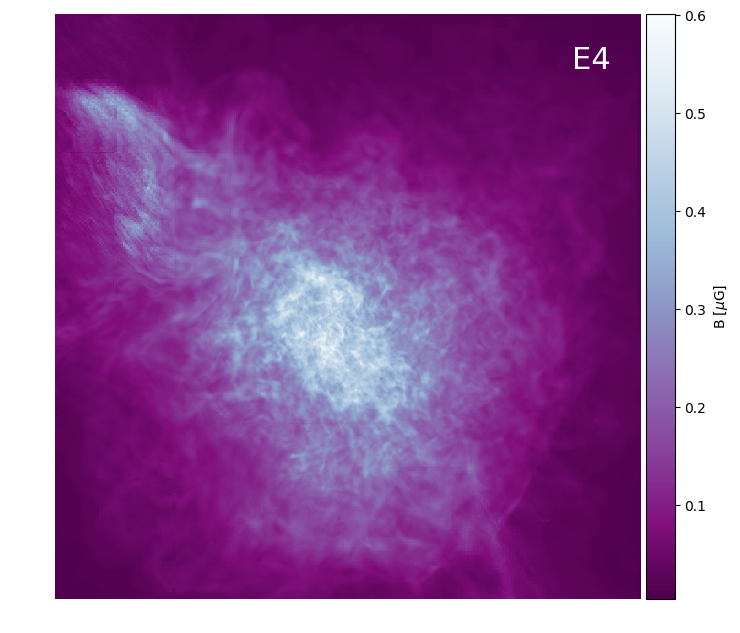}
    \caption{\footnotesize \textit{Left:} Magnetic energy spectra of all of our cluster sample at $z=0$. The solid lines
    correspond to the data and the scatter plots show our best-fit. \textit{Right:} Projection of the magnetic field strength in
    galaxy cluster E4 at $\Delta x_{\rm max}$}
    \label{fig01}
\end{figure}

 
 We analyse clusters with different
dynamical states at redshift $z=0$, namely clusters with ongoing mergers,
relaxed ones and clusters that have suffered a recent major merger.
 The baseline model we tested with our cosmological simulation relies on an analytic solution 
 for the magnetic power spectrum derived from dynamo theory (\cite{Kazan,Kul}).
 Our tests (Dom\'inguez-Fern\'andez et al., to be submitted) show that all spectra
 can be fitted well by an equation which is directly derived from dynamo theory
 (see Eq. in Fig.~\ref{fig01}).  In Fig.\ref{fig01} we show the best-fit of the magnetic energy spectra
of all of the clusters in our sample.

\begin{discussion}

1) Regardless of the dynamical state of each cluster, the magnetic energy spectral shape is in good agreement
with dynamo theory.  As a first approximation,  this indicates that this model can be used for a detailed analysis of
a cluster evolution where a small-scale dynamo can be  
acting in addition to gas compression and shocks.

2) The normalization of the magnetic energy spectrum overall is determined by the dynamical state of each cluster.
    Relaxed clusters have the highest value for the magnetic spectrum, followed by post-mergers clusters, and then merging clusters. This result is 
    tentatively consistent with the fact that in relaxed system a small-scale 
    dynamo would have to be active for longer dynamical times. Nevertheless, we do not observe a clear dependence on the
    virial mass of each cluster due to the small number of clusters in our sample.
    
3) Our analysis shows that the magnetic power spectra does not retain information on the last major-merger,
as minor mergers are also injecting turbulence during the formation of each cluster.

4) We refer the reader to the complete work (Dom\'inguez-Fern\'andez et al., to be submitted) where a detailed analysis on the cluster sample is done and
where also the evolution of a merging cluster is studied.

5) The resulting power spectrum shows that in general the three-dimensional components of
the magnetic field are non-Gaussian. This results from having different magnetic field components being continuously injected by the accretion of substructures (\cite{Vazza18}) 
rather than from the presence of highly intermittent MHD turbulence (e.g. \cite{Shuk}). This has direct consequences for Faraday Rotation measurements and illustrates 
the importance of reliable simulations of magnetic fields in galaxy clusters for an accurate modelling of future polarisation surveys 
(e.g. \cite{Johnson} ).

\end{discussion}

\section*{\footnotesize Acknowledgements}
{\scriptsize Our simulations were performed using {\enzo} and 
the Supercomputing resources at the Juelich Supercomputing Centre (JSC), under project HHH42. 
We  acknowledge the European Union's Horizon 2020 program under the ERC Starting Grant "MAGCOW", no. 714196. }

\end{document}